# Pure thermal spin current and perfect spin-filtering with negative differential thermoelectric resistance induced by proximity effect in graphene/silicene junctions

Zainab Gholami and Farhad Khoeini*

*Department of physics, University of Zanjan, P.O. Box 45195-313, Zanjan, Iran*

**Abstract**

The spin-dependent Seebeck effect (SDSE) and thermal spin-filtering effect (SFE) are now considered as the essential aspects of the spin caloritronics, which can efficiently explore the relationships between the spin and heat transport in the materials. However, there is still a challenge to get a thermally-induced spin current with no thermal electron current. This paper aims to numerically investigate the spin-dependent transport properties in hybrid graphene/silicene nanoribbons (GSNRs), using the nonequilibrium Green's function method. The effects of temperature gradient between the left and right leads, the ferromagnetic exchange field, and the local external electric fields are also included. The results showed that the spin-up and spin-down currents are produced and flow in opposite directions with almost equal magnitudes. This evidently shows that the carrier transport is dominated by the thermal spin current, whereas the thermal electron current is almost disappeared. A pure thermal spin current with the finite threshold temperatures can be obtained by modulating the temperature, and a negative differential thermoelectric resistance is obtained for the thermal electron current. A nearly zero charge thermopower is also obtained, which further demonstrates the emergence of the SDSE. The response of the hybrid system is then varied by changing the magnitudes of the ferromagnetic exchange field and local external electric fields. Thus, a nearly perfect SFE can be observed at room temperature, whereas the spin polarization efficiency is reached up to 99%. It is believed that the results obtained from this study can be useful to well understand the inspiring thermospin phenomena, and to enhance the spin caloritronics material with lower energy consumption.

## 1. Introduction

The spin caloritronics has now constructed the subject of many researches in recent years [1-3]. Spin caloritonics combines the spintronics [4-6] and thermoelectrics to study the interactions among spin and charge in the presence of temperature bias. It can also provide many different approaches for the thermoelectric waste heat recovery, future information, and device technologies [7-9]. The spin caloritronics usually show different innovative effects such as the SDSE [10], spin Seebeck diode (SSD) effect [11], thermal SFE [12], and the thermal giant magnetoresistance phenomenon [13]. According to the experimental results, the spin Seebeck effect (SSE) initially showed an interaction between the spin and heat currents in the magnetic metals, simultaneously [14]. Thus, there is SSE in different magnetic phases such as ferromagnetic metals, semiconductors, and insulators [15-17], paramagnetic and antiferromagnetic materials [18,19], and nonmagnetic materials if a magnetic field is applied [20]. It is worth mentioning that SDSE and SSE phenomena are both induced due to the interaction between the spin and heat currents. However, the magnons are the carriers in SSE, whereas they are the electrons in SDSE.

In SDSE, the temperature bias can generate a spin current, as an essential physical quantity in spintronics, and provide an efficient way for controlling the electron spins in the presence of the temperature gradient. Since the electron current is often accompanied by Joule heating, the spin current in the low-power-consumption nanodevices is required to be used while the electron current is decreased as much as possible. This is now achievable by SDSE. As shown in the previous studies [21], the spin current may be produced by the opposite current direction for different spins. An almost non-dissipative SDSE can be obtained if the electrons in two spin channels, spin-up and spin-down flow in opposite directions with equal values, which is called the perfect SDSE [22]. It is noted that an essential factor in producing the spin current using a temperature gradient is to find a suitable spin-thermoelectric material, which can keep the spin thermopower with opposite signs for different spins [23].

The Graphene (GE), as the first isolated two-dimensional (2D) material, is composed of a honeycomb structure of carbon atoms, and have widely become popular among researchers. GE often shows different thermal [24,25], mechanical [26,27] and optical [28] properties. However, GE has uncommon transport properties, which are related to its unusual electronic structure [29,30]. This is especially true around the Fermi level, where the charge carriers behave similar to massless particles. Because of the long spin relaxation time and length properties of the GE, it is identified as a promising candidate for the future nanoelectronic and spintronic applications [31]. The magnetic and thermoelectric properties of GE have also attracted great interest in recent years. Some of the applications of this material have already been proposed and discussed in [32,33].

Owing to successful studies of GE, some serious attempts have been made by researchers to develop new forms of low-dimensional materials. More recently, Silicene (SE), a hexagonal atomic structure with silicon atoms two-dimensionally bonded together, was first theoretically developed in the literature [34]. SE was then fabricated by depositing Si on Ag [35], and Ir [36] surfaces in the laboratory environment. The structural stability of the SE was then confirmed by other researchers using the phonon spectra calculations [37]. SE and GE have similar electronic structures near the Fermi level and hence can result in the massless Dirac fermions [38]. This concept is now widely used for developing the high-performance field-effect transistors [39]. One of the properties of SE is that it has a larger bandgap, induced by the spin-orbit interaction (SOI). This establishes the quantum spin Hall effect [40], and has a significant role in spin transport and spintronic devices. The bandgap in SE can be opened and controlled by applying an external electric field normal to the atomic plane [41]. However, GE has not such a property. There is also the interaction between the electromagnetic field and spin-orbit coupling in the SE, a feature that can be used to study physics in quantum phase transition [42]. More recently, several studies have been carried out by researchers to investigate the charge and spin thermal transport properties of the SE [43-45].

As shown in the previous studies (e.g., [46]), the SDSE can emerge in the graphene-based nanodevices. However, because GE shows an extremely low figure of merit [47], it was introduced as an inefficient thermoelectric material in the next researches [48]. Hence, GE rarely shows a perfect SDSE. On the other hand, the SE may be a more promising material than GE in thermoelectrics. The previous works showed that SE could significantly improve the Seebeck coefficient [49] due to its nonzero energy gap. SE also has some special properties in thermoelectric. For instance, the spin-dependent thermoelectric transport properties of the zigzag SE nanoribbons (ZSNRs) have been studied. The results showed that these ZSNRs could show a high spin-filter efficiency [50]. Another type of ZSNRs was then studied [10] and a perfect SDSE has been observed. The theoretical studies showed that the quasi-one-dimensional wire can provide more significant thermoelectric properties than those of 2D structures [51]. This may be related to the significant changes in the electronic and thermal properties of the material. It is worth mentioning that the thermoelectric properties of the nanostructures can be further improved by different modifications such as hybridization [52], doping [53], absorption [54], and so on. The past theoretical studies showed that the hybrid nanostructures such as hybrid $MoS_2/WS_2$ [55] and BN/GE nanoribbons [56] have higher thermoelectric properties than single nanostructures. In recent years, lateral and vertical GE/SE heterojunctions have theoretically [57-59] and experimentally [60,61] been studied. The results showed that the hybrid systems composed of GE and a different 2D material [62,63] (e.g., SE) could remarkably help to produce a structural type with different properties and applications. However, there is now lack of data about the spin transport, electronic behavior, and thermoelectric properties of the hybrid GSNRs, and further studies are still required in this field.

In this paper, the electronic and thermal spin-dependent transport properties of the hybrid GSNRs are numerically investigated. Besides, in this study, the tight-binding (TB) approach based on the nonequilibrium Green's function (NEGF) method is used. The effects of ferromagnetic exchange field, together with the local external electric fields, are also included. A practical way is then proposed to achieve the SDSE, SFE, and negative differential thermoelectric resistance (NDTR) in the hybrid GSNRs. The main findings of the current study are reported and discussed in two separate sections, including (1) SDSE and (2) SFE. In the SDSE case, the spin-up and spin-down currents with almost equal magnitudes are produced in opposite directions (i.e., nearly perfect SDSE) if a temperature gradient is created between the left and right leads. It is noted that because the spin-polarized conduction electrons are the carriers, thus the results reported in this research are limited to SDSE the phenomenon. A nearly pure spin with very small charge thermopower is also observed. However, the spin polarization has reached up to 99% in the SFE case. It is noted that because of the competition between the spin-dependent currents, some interesting transport features such as the change of the flowing direction and NDTR is observed. This evidently confirms its potential thermoelectric device applications by selecting various device temperature sets.

## 2. Models and theoretical calculations

## 2.1. Device structure

The SE atoms have an in-plane distance of 2.30 Å, which is much larger than the bond length of GE, i.e., 1.42 Å. In addition, the next-nearest-neighbor distance in the GE is equal to 2.46 Å, which is almost 7% larger than the in-plane atomic distance of SE. Thus, the construction of a lattice-matched one-dimensional interface between the GE and SE is impossible along the same chirality [58]. On the other hand, there is a small out-of-plane buckling in the SE, and thus it is not located in a plane [42]. To consider this buckling feature, the Si atoms in the GE/SE nanocomposite are first oriented in an out-of-plane mode with two parallel sublattices $A$ and $B$ vertically separated by a distance of 0.46 Å. The interface is then established via extending the SE by 7% along the $y$ direction to match the lattice of GE. The distance between the C and Si atoms at the interface is equal to 1.80 Å [57]. Figure 1 shows the atomic configuration of the hybrid GSNR selected for this study. As shown in figure 1, the hybrid GSNR is constructed using three different parts, including the semi-infinite metallic armchair GE nanoribbons (AGNRs) on the left and right electrodes and a ZSNR in the central region [64]. Some dangling carbon bonds also remain at the GE/SE interface, which can be passivated by an additional hydrogen atom. It is assumed that the length of central region is almost equal to $M=8$ unit cells. Each unit cell is composed of $N_S=16$ atoms, leading to a length and width of 29.85 and 25.27 Å, respectively, for the central region. Besides, each unit cell of the left and right leads includes $N_G=46$ atoms.

## 2.2. Tight-binding and Green's function method

In this subsection, the electronic quantum transport in the quasi-one-dimensional hybrid GSNR structure is studied. As explained earlier, the hybrid GSNR system is in the $xy$ plane, and consists of three different regions, including the AGNRs on the left and right leads and a ZSNR in the middle scattering region (see figure 1(a)). The generalized Hamiltonian for the system can be described as follows:

$$H_T = H_L + H_R + H_C + H_{CL} + H_{CR}, \qquad (1)$$

in which, $H_{R(L)}$ and $H_C$ represent the Hamiltonian for the isolated right (left) lead and central region, respectively. Using the TB approximation model [40,42,65], the Hamiltonian of the right (left) lead and the central region are obtained as follows:

$$H_{R(L)} = -t_{R(L)} \sum_{<i,j>,\alpha} c_{i\alpha}^\dagger c_{j\alpha} + eE_{yG} \sum_{i,\alpha} y_{iG} c_{i\alpha}^\dagger c_{i\alpha} + \text{H.c.}, \qquad (2)$$

$$H_C = -t_C \sum_{<i,j>,\alpha} c_{i\alpha}^\dagger c_{j\alpha} + i\frac{\lambda_{so}}{3\sqrt{3}} \sum_{\ll i,j\gg,\alpha,\beta} v_{ij} c_{i\alpha}^\dagger (\sigma_z)_{\alpha\beta} c_{j\beta} + M_z \sum_{i,\alpha} c_{i\alpha}^\dagger \sigma_z c_{i\alpha} + \qquad (3)$$

$$+ elE_z \sum_{i,\alpha} \xi_i c_{i\alpha}^\dagger c_{i\alpha} + eE_{yS} \sum_{i,\alpha} y_{iS} c_{i\alpha}^\dagger c_{i\alpha} + \text{H.c.},$$

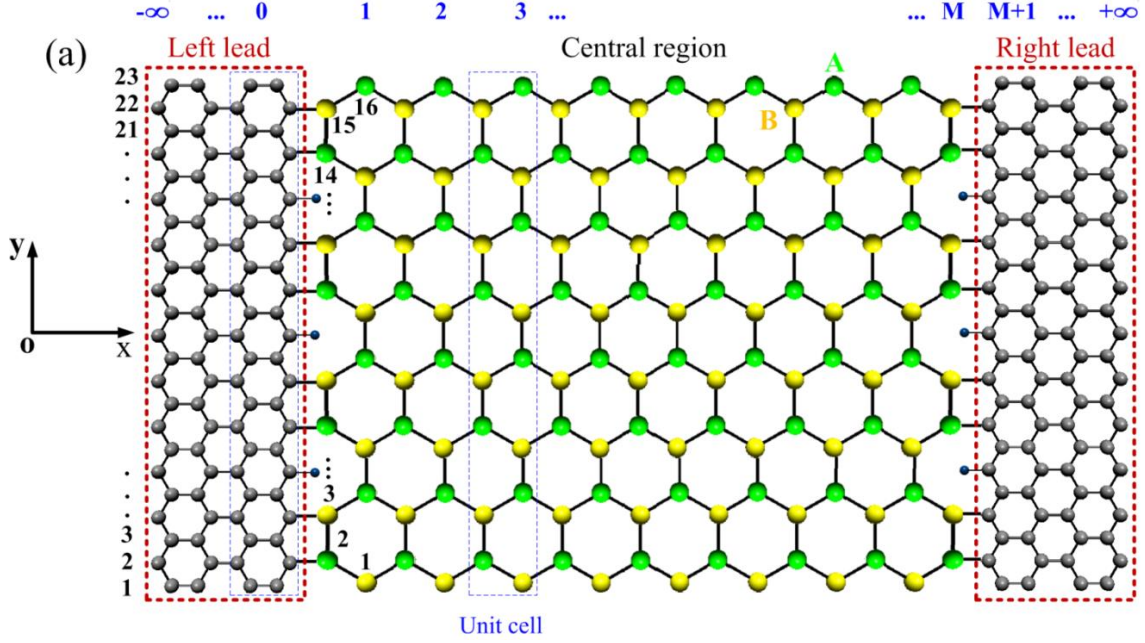

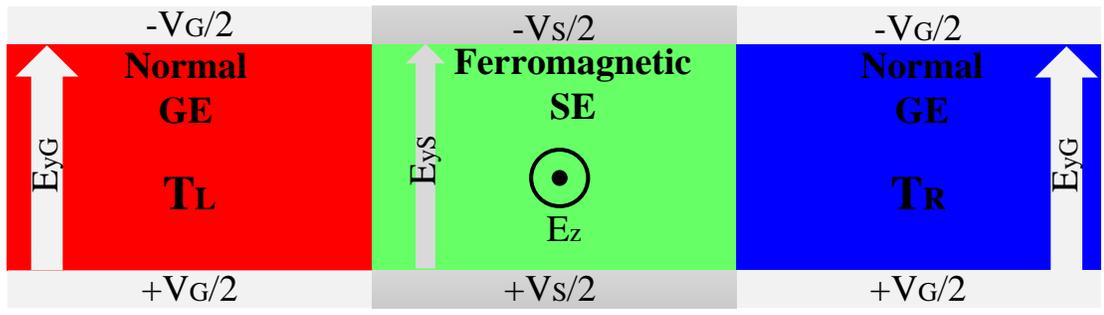

**Figure 1.** (a) A schematic representation of the hybrid graphene/silicene nanoribbon (GSNR) adopted for this study, the central region consists of about $M = 8$ unit cells, each has $N_S = 16$ atoms; each unit cell of the leads also has $N_G = 46$ atoms, (b) the external electric and ferromagnetic exchange fields applied to the system, where $T_L$ and $T_R$ define the temperature of the left and right leads, respectively.

where $t_{R(L)}$ and $t_C$ are the hopping energies between the nearest-neighbor atoms in the right (left) lead, and the central region, so their values are equal to 2.66 and 1.60 eV, respectively. $c_{i\alpha}^{\dagger}(c_{i\alpha})$ creates (annihilates) an electron with spin $\alpha$ at atom $i$. $<i,j>$ and $<<i,j>>$ represent the nearest-neighbor and next-nearest-neighbor pairs, respectively. $\lambda_{so}$ is the effective SOI parameter and is equal to 3.9 meV for the central region [42]. Due to the weak SOI in the GE [66], the values of $\lambda_{so}$ and $\lambda_r$ are assumed equal to zero for the left and right leads. $\boldsymbol{\sigma} = (\sigma_x, \sigma_y, \sigma_z)$ is the Pauli matrix. $v_{ij}$ is defined as $v_{ij} = (\boldsymbol{d}_i \times \boldsymbol{d}_j)/|\boldsymbol{d}_i \times \boldsymbol{d}_j|$ where $\boldsymbol{d}_i$ and $\boldsymbol{d}_j$ are the two nearest bonds connecting the next-nearest-neighbors. $\xi_i = +1, -1$ for the sublattices $A$ and $B$, respectively; and $2l$ is the buckling distance for the SE. $E_z$ is also the perpendicular electric field and produces a voltage difference of $2elE_z$ between the two sublattices, where $e$ is the electron charge.

This causes the electrons to experience a staggered potential when jumping from a site to its nearest-neighbor. It is worth mentioning that the external electric field can efficiently differentiate the SE from GE. $M_z$ is the ferromagnetic exchange field, which can be produced by the proximity effect of a ferromagnetic material [42]. $E_{yG}$ and $E_{yS}$ are the inhomogeneous transverse electric field components along the *x* direction, applied to the leads and central region, respectively. $y_{iG}$ is also the normal distance of atom *i* from the middle of the ribbon.

$H_{CR(CL)}$ also shows the Hamiltonian for the coupling between the central region with the right (left) lead and using the TB approximation can be written as

$$H_{CR(CL)} = -t_{CR(CL)} \sum_{<i,j>,\alpha} c_{i\alpha}^{\dagger} c_{j\alpha} + \text{H.c.,} \qquad (4)$$

where $t_{CR(CL)}$ represents the hopping energy between the central region and right (left) lead. The contact hopping energy can also be determined by averaging the values obtained from the Harrison's scaling law [67,68]. In this study, namely, the hopping energies are re-calculated for the GE and SE at the interface when the C and Si atoms bond-lengths are changed to 1.8 Å. The geometric mean of the obtained hopping energy values (i.e., 2.07 eV) will then be used as the $t_{CR(CL)}$ parameter, which is required to be the same order of $t_{R(L)}$ and $t_C$ [55]. The electronic transport is assumed ballistic. Such an assumption is valid when the mean free path of the carriers, which is in the order of the micron in the SE and GE at room temperature, is larger than the sample dimensions. This is due to the high mobility of charge carriers in the SE and GE, and thus they can simply move long distances without inelastic scattering. In the central region, the spin direction of the electron is assumed preserved, and the spin-flip scattering is ignored. Thus, the spin-up and spin-down electron transports can individually be studied. This assumption is correct because the spin diffusion length in the SE is in the order of several micrometers [29,69].

To determine the spin-dependent current, it is first required the spin-dependent electron transmission function to be calculated. The Green's function of the central region can be written as follows [70]:

$$G_{C,\alpha} = \left[ (E + i0^+)I - H_C - \Sigma_{L,\alpha} - \Sigma_{R,\alpha} \right]^{-1}, \qquad (5)$$

where $\Sigma_{R(L),\alpha}$ is the right (left) self-energy matrix and is computed by Eq. (6). $\Sigma_{R(L),\alpha}$ includes the effect of two semi-infinite AGNRs on the central region.

$$\Sigma_{R,\alpha} = H_{CR}\, g_{R,\alpha}(E)\, H_{CR}^{\dagger}, \qquad (6a)$$

$$\Sigma_{L,\alpha} = H_{LC}^{\dagger}\, g_{L,\alpha}(E)\, H_{LC}, \qquad (6b)$$

in which, $g_{R(L),\alpha}$ is the surface Green's function of the right (left) lead and is computed by

$$g_{R(L),\alpha}(E) = \left[ (E + i0^+)I - H_{R(L)}^{00} - H_{R(L)}^{01} T_{R(L)} \right]^{-1}, \qquad (7)$$

where $H_{R(L)}^{00}$ defines the TB Hamiltonian matrix of the unit cell studied in this research for the right (left) lead. $H_{R(L)}^{01}$ is the hopping matrix between two adjacent unit cells in the right (left) lead. $T_{R(L)}$ is the transfer matrix of the right (left) lead and can be determined using an iterative methodology, as proposed in Reference [70]. The electron transmission function can then be obtained using the following equations [70]:

$$T_\alpha(E) = Tr[\Gamma_{L,\alpha}(E) G_{C,\alpha}(E) \Gamma_{R,\alpha}(E) G_{C,\alpha}^\dagger(E)], \tag{8}$$

where $\Gamma_{R(L),\alpha}$ is the broadening function and describes the coupling between the right (left) lead with the central region. $\Gamma_{R(L),\alpha}$ is given by

$$\Gamma_{R(L),\alpha} = i(\Sigma_{R(L),\alpha} - \Sigma_{R(L),\alpha}^\dagger). \tag{9}$$

### 2.3. Spin-dependent thermoelectric

The thermally-induced spin-dependent electric current can be computed by using the Landauer-Büttiker expression, as follows [71]:

$$I_\alpha = \frac{e}{h} \int_{-\infty}^{+\infty} T_\alpha(E)[f_L(E,T_L) - f_R(E,T_R)]\, dE, \tag{10}$$

where $f_{R(L)}$ is the Fermi-Dirac distribution. $e$, $h$, and $T_{R(L)}$ are also the electron charge, Plank constant, and temperature of the right (left) lead, respectively. As shown in Eq. (10), the temperature gradient $\Delta T = T_L - T_R$ between the leads produces a nonzero value of $f_L - f_R$, and thus the spin-dependent current is simply identified as a function of $\Delta T$ and $T_{R(L)}$. The net spin and charge currents can then be obtained by $I_S = I_{up} - I_{dn}$ and $I_C = I_{up} + I_{dn}$, respectively. To compute the other spin-dependent thermoelectric quantities, an intermediate function can be defined as follows [71]:

$$L_{n,\alpha}(E_F, T) = -\frac{1}{h} \int (E - E_F)^n \frac{\partial f(E, E_F, T)}{\partial E} T_\alpha(E) dE, \tag{11}$$

where $E_F$ and $T$ are the Fermi Energy and temperature values, respectively. Assuming a linear response regime, $\Delta T \cong 0$, the other spin-dependent thermoelectric parameters such as the spin-dependent thermopower, electrical conductance, and the electron's contribution to the thermal conductance can be calculated [72]. Based on Eq. (11), the spin-dependent Seebeck coefficient can then be calculated as follows:

$$S_\alpha = -\frac{1}{eT}\left(\frac{L_{1\alpha}}{L_{0\alpha}}\right). \tag{12}$$

where $S_S = S_{up} - S_{dn}$ and $S_C = (S_{up} + S_{dn})/2$ are the spin and charge Seebeck coefficients, respectively [72].

### 3. Results and discussion

In this section, the results of the current study are presented and discussed. For this purpose, Eqs. (8) and (10), along with the intermediate function, have been used to evaluate the obtained results. A hybrid nanoribbon system is first defined and then subjected to different local external fields, including the perpendicular electric ($E_Z$) and ferromagnetic exchange ($M_z$) fields, which are applied to the central region (see figure 1). The application of the local external electric [73] and ferromagnetic exchange fields are now experimentally feasible. For example, a ferromagnetic exchange field can be created by the proximity with a ferromagnetic insulator EuO as suggested for GE [74]. An inhomogeneous transverse electric field with the magnitude of $E_{yS}$ and $E_{yG}$ are also applied to the device, the left and right leads, respectively [75]. The electric current obtained by the temperature difference ($\Delta T$) without any external bias voltage, i.e., the difference between the temperature of the left, $T_L$, and the right, $T_R$, leads are calculated.

**3.1. Spin-dependent Seebeck effect**

In order to study the thermal spin transport properties of the considered hybrid GSNRs, the values of the spin-up ($I_{up}$) and spin-down ($I_{dn}$) currents are determined as a function of $T_R$ for three different temperature gradients $\Delta T$=10, 20, and 40K (see figure 2). The values of $M_z$ = 0.181 eV, $E_z$ = 0.081 V/Å, $E_{yS}$ = 0.127 V/Å, and $E_{yG}$ = 0.913 V/Å are selected, and a positive spin-up and negative spin-down currents are evidently observed. These values have been selected based on the previous studies (e.g., [76-79]), and it is believed that they can provide relatively large symmetric spin currents. A nearly perfect SDSE is identified in this hybrid GSNRs [14,15,80]. Because the spin-up and spin-down currents have only produced due to the temperature gradient, and they flow in opposite directions with almost equal magnitudes. The GSNR evidently shows a good insulating behavior without any charge or spin current for the low-temperature values. As shown in figure 2(a), the threshold temperatures of $T_{th}$ = 60, 50, and 40 K are almost obtained for $\Delta T$=10, 20, and 40K, respectively; where the spin-up and spin-down currents are almost zero for $T_R < T_{th}$, whereas the spin currents proportionally increase with respect to $T_R$ for $T_R > T_{th}$. The spin currents also increase in opposite directions as $\Delta T$ increases (see figure 2(b)) such that they vary linearly and symmetrically respect to the zero-current axis for the entire range of $\Delta T$.

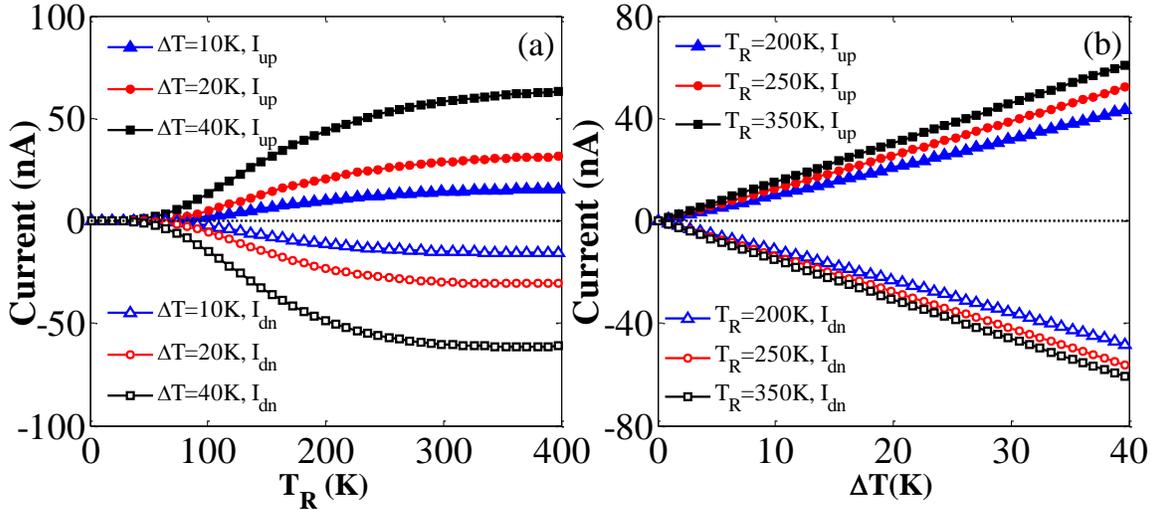

**Figure 2.** The thermally-induced spin currents for the considered hybrid GSNRs (a) the variation of the spin currents versus the right lead temperature, $T_R$, for different temperature gradients, $\Delta T$= 10, 20, and 40 K. The spin-up ($I_{up}$) and spin-down ($I_{dn}$) currents flow in positive and negative directions, respectively (i.e., SSE) (b) the variation of the spin currents versus $\Delta T$ for the right lead temperatures $T_R$=200, 250, and 350 K.

The SDSE is further confirmed by all these issues. It is worth mentioning that the studied hybrid GSNRs can significantly provide larger thermally-induced spin currents compared to those given by the pristine AGNRs and ZSNRs.

In order to well understand the SDSE phenomenon in the considered hybrid GSNRs, the Landauer Büttiker formula (Eq. 10) is studied in detail. As can be seen from Eq. (10), the transport coefficients and the difference between the Fermi-Dirac distributions of the left ($f_L$) and right ($f_R$) electrodes provide the two factors affecting the thermal spin-dependent currents. Because the leads are composed of similar material and density of states, the difference in carrier concentrations between these two leads is due to the temperature difference and is calculated using the Fermi-Dirac distribution. It is noted that $f_L - f_R$ is an odd function around the Fermi level, and thus the sign of the current function is determined by the slope of the transmission coefficients for this region. $f_L - f_R$ is almost equal to zero for the higher energy values (see the inset of figure 3(d)) [10]. As the carriers with the energies higher (lower) than the Fermi level flow from the left (right) lead towards the right (left) lead, the electron, $I_e < 0$ (hole, $I_h > 0$) currents are generated, respectively. A net zero thermal current is obtained when the transmission spectrum is symmetric, and $I_e$ and $I_h$ currents neutralize each other. The variations of the spin-dependent transmission coefficients ($T_{up}$ and $T_{dn}$) versus the energy $E-E_F$ around the Fermi level (here $E_F$ is set to zero) are shown in figure 3(a). As shown in figure 3(a), because the spin-up and spin-down transmission spectra have been located below and above the Fermi level, $I_{up}$ and $I_{dn}$ will also have opposite signs.

Two narrow transmission bands for the spin-up and spin-down electrons almost occur within the range of -0.2 eV < $E-E_F$ < -0.04 eV and 0.04 eV < $E-E_F$ < 0.2 eV, respectively (see figure 3(a)). It is worth mentioning that the electron-hole symmetry is often disturbed by these transmission peaks and result in a nonzero net spin current [81]. For further explanation, the peak value of the spin-down electron transmission has been located above the Fermi level, and thus the electrons can be transported from the right lead to the left one. This leads to a negative spin-down current. On the contrary, because the peak value of the spin-up electron transmission occurs below the Fermi level, the transport of holes produces a positive spin-up current from the left lead to the right one (see figure 3(a)). There also exist two peak values for the spin-up and spin-down electrons in the above and below the Fermi level. However, compared to the former peak values, they are very small. Accordingly, a nearly perfect SDSE is observed in the hybrid GSNRs, because the transmission peaks for the spin-up and spin-down electrons are almost symmetric relative to the Fermi level. In this study, a relatively moderate threshold temperature $T_{th}$ (~50 K) for each spin current is needed to extend the Fermi distribution and cover the transmission peaks. This causes a nonzero spin current to be produced between the left and right leads. All these issues may be attributed to the fact that the Fermi distribution is exponentially decreased, and the transmission spectrum shows a relatively medium energy gap for the spin-up and spin-down electrons.

To assess the combined effects of external fields on the hybrid GSNRs, the band structures of the leads and central region are also shown in figure 3(b)-(c). The lowest-energy subbands for each band structure are related to the electron and hole for $E-E_F > 0$ and $E-E_F < 0$, respectively. As shown in figure 3(b), the spin-dependent subbands of the leads are almost matched, whereas the lowest-energy subbands of the central region belong to two different spin states (see figure 3(c)). The effects of each external field are (1) the ferromagnetic exchange field ($M_z$ = 0.181 eV) causes different spins to shift in opposite directions and lead to a huge spin splitting [74]. In other words, the proximity effect causes the gapless spin edge states are damaged (2) the semiconducting gaps are also generated between the hole and electron subbands when a perpendicular electric field ($E_z$=0.081 V/Å) is applied [41]. This is due to the fact that the inversion symmetry is broken by the staggered sublattice potential (3) the transverse electric field ($E_{yS}$= 0.127 V/Å and $E_{yG}$= 0.913 V/Å) produces the gaps between the hole and electron subbands, and causes the electron states to shift to the lower energies, whereas the hole states shift to the higher ones [82]. The band structure shown in figure 3(b)-(c) is obtained when all these fields are applied simultaneously.

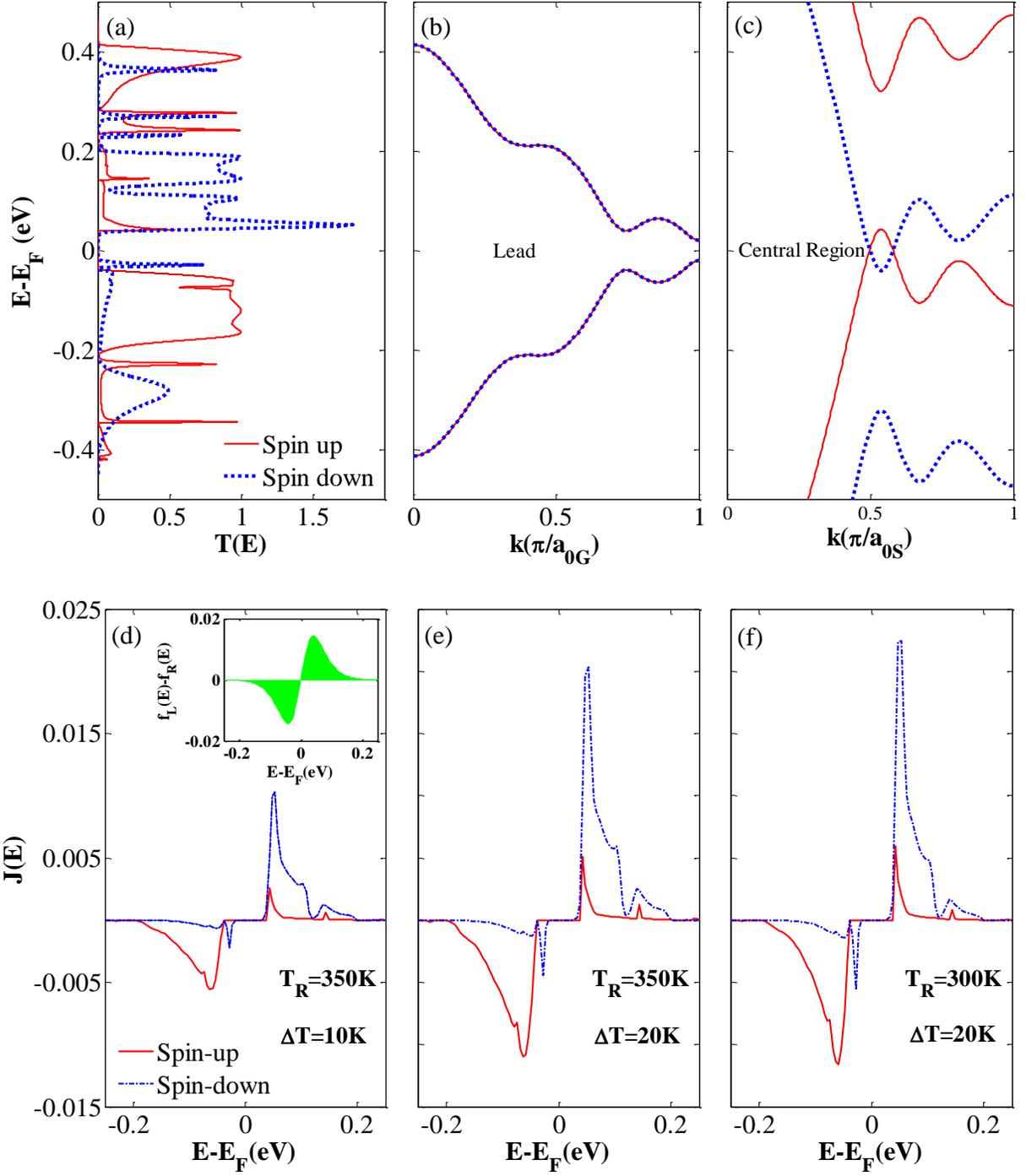

**Figure 3.** (a) The spin-dependent transmission spectra ($T_{up}$ and $T_{dn}$) versus $E-E_F$, (b) the band structures of the right (left) lead with an applied electric field of $E_{yG}$= 0.913 V/Å, (c) the band structures of the ferromagnetic central region including an exchange field of $M_z$= 0.181 eV, electric fields of $E_Z$= 0.081 V/Å and $E_{yS}$= 0.127 V/Å, and the variation of the spin-dependent current spectrum $J(E)$ versus $E-E_F$ for (d) $T_R$ = 350 and $\Delta T$ = 10 K, (inset shows the difference between the Fermi–Dirac distributions for the left and right leads, $f_L(E)$-$f_R(E)$, as a function of $E-E_F$), (e) $T_R$ = 350 and $\Delta T$ = 20 K, and (f) $T_R$ = 300 and $\Delta T$ = 20 K.

Figures 3(d) to 3(f) show the current spectrum $J=T(f_L-f_R)$ of the spin currents, plotted for different temperature sets. The area limited to the current spectrum curve and the energy axis ($E$-$E_F$) defines the spin current (i.e., $I_{up}$ and $I_{dn}$). As shown in figures 3(d) to 3(f), the spin-up current spectra are relatively symmetric with respect to the spin-down ones about the Fermi level and with almost equal areas for the entire different temperature sets. A comparison between the exact areas obtained from the current spectrum ($J$) for the spin-up and spin-down clearly shows that there is a very small difference between the spin-up and spin-down areas in each plot. This again confirms the nearly perfect SDSE. For $I_{up}$, the peak value of the current spectrum at $\Delta T$= 20 K is larger than that of $\Delta T$= 10 K when $T_R$ is assumed equal to 350 K. This clearly shows that the spin current grows as $\Delta T$ increases. Nonetheless, the area under the $J$ curves for $T_R$ =300K is smaller than that of $T_R$ =350K when $\Delta T$ is assumed to be 20K. This indicates that the spin currents grow with the increase of $T_R$. Because, there is a bandgap in the transmission spectrum, a zero $J$ value is also observed around the Fermi level.

The thermal-driven net spin currents, $I_S$ (= $I_{up} - I_{dn}$), and the total charge currents, $I_C$ (= $I_{up} + I_{dn}$), are also calculated. In figure 4, $I_S$ and $I_C$ are plotted versus $T_R$ and $\Delta T$. As shown in figure 4(a) and 4(c), $I_S$ value is generally increased as $T_R$ or $\Delta T$ values are increased. For example, $I_S$ value is almost 70, 86, and 100 times larger than those of $I_C$ for $T_R$=400 K and $\Delta T$=10, 20, and 40 K, respectively, which are much greater than those reported in [10,83]. Thus, the carrier transport through the hybrid nanostructure is controlled by the spin current. $I_C$ also shows some interesting transport properties when $T_R$ or $\Delta T$ increases. For example, $I_C$ is zero for $T_R < T_{th}$ when $\Delta T$ is equal to 20 K (see figure 4(c)). $I_C$ also drops to the negative values when $T_R$ is larger than $T_{th}$. This indicates the appearance of a thermoelectric switcher. As $T_R$ is further increased, $I_C$ reaches to its peak value where the negative differential thermal resistance emerges [84]. Indeed, the NDTR occurs due to the competition between $I_{up}$ and $I_{dn}$ with opposite flowing directions. $I_C$ also decreases to a zero value as $T_R$ increases to a critical temperature value of $T_R$ =358 K. This clearly confirms the emergence of the thermal-induced pure spin current. The flowing direction of $I_C$ changes for $T_R > 358$ K, because its sign gets reverse. The variations of $I_C$ versus $\Delta T$ are also computed and plotted in figure 4(d). As shown in figure 4(d), $I_C$ is negative for the smaller values of $T_R$ =200 and 250 K, and different values of $\Delta T$, whereas $I_C$ is almost equal to zero for the relatively large values of $T_R$ =350 K. The observed behaviors clearly confirm that the current of the hybrid GSNRs is appropriate for different device applications by selecting various device temperature sets.

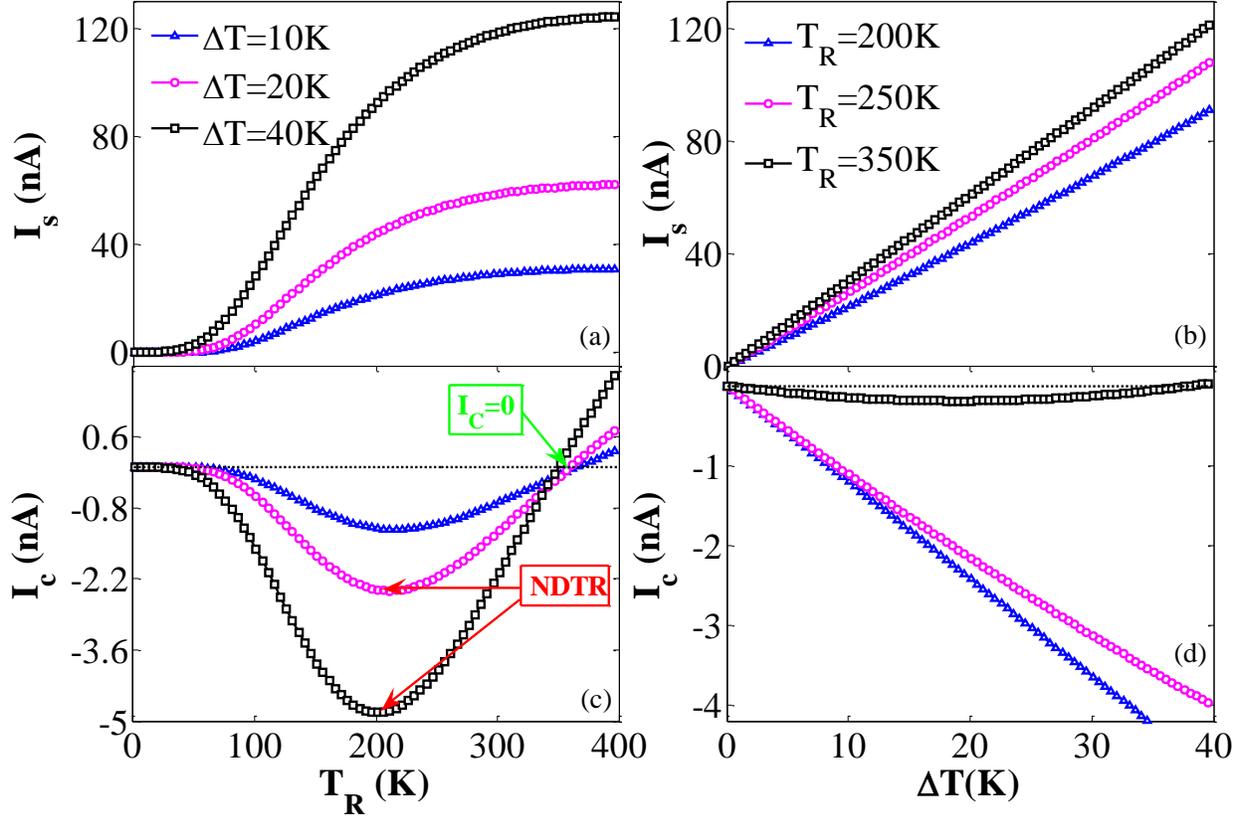

**Figure 4.** (a) The variation of the net spin current ($I_S = I_{up} - I_{dn}$) versus $T_R$ for $\Delta T$=10, 20, and 40 K, (b) the variation of $I_S$ versus $\Delta T$ for $T_R$ = 200, 250, and 350 K, (c) the total electron current ($I_C = I_{up} + I_{dn}$) as a function of $T_R$ for $\Delta T$= 10, 20, and 40 K, and (d) the variation of $I_C$ versus $\Delta T$ for $T_R$ = 200, 250, and 350 K.

### 3.2. Thermal spin-filtering effect

The thermal SFE without any threshold temperatures in the hybrid GSNRs is studied in this subsection. The effects of an electric field of $E_z = 0.030$ V/Å and an exchange field of $M_z = 0.157$ eV, applied perpendicularly to the central region of the hybrid GSNR are also investigated. The inhomogeneous transverse electric fields equal to $E_{yS} = 0.575$ V/Å and $E_{yG} = 0.029$ V/Å are also considered. Figure 5(a) shows the variations of the thermally-induced currents versus $T_R$ for different values of $\Delta T$. As shown in figure 5(a), larger values of $I_{up}$ are obtained for the high temperatures, whereas $I_{dn}$ is zero for the entire range of temperature as $T_R$ increases. The nearly perfect thermal SFE is valid [11,12,83], and this evidently illustrates that the spin-up transport channels are opened, whereas the spin-down transport channels always remain close. For example, $I_{up}$ reaches to its maximum value and is then decreased to zero at a critical temperature value of $T_R$ =172 K and $\Delta T$=20 K. However, $I_{up}$ sign is reverse and the flowing direction is changed for $T_R$ > 172 K. It is noted that when the $I_{up}$ is maximum, the NDTR also reaches its peak value. Again, the NDTR occurs for the larger values of temperature and $\Delta T$. As a result, the hybrid GSNR can be used as a thermal spin device with various multiple attributes. Figure 5(b) also shows the variations of $I_{up}$ and $I_{dn}$ versus $\Delta T$ for different values of $T_R$=200, 250, and 350 K. As shown in figure 5(b), $I_{up}$ increases when $T_R$ and $\Delta T$ are increased, whereas $I_{dn}$ remains almost zero for different values of $\Delta T$. This further confirms that the SFE has been produced.

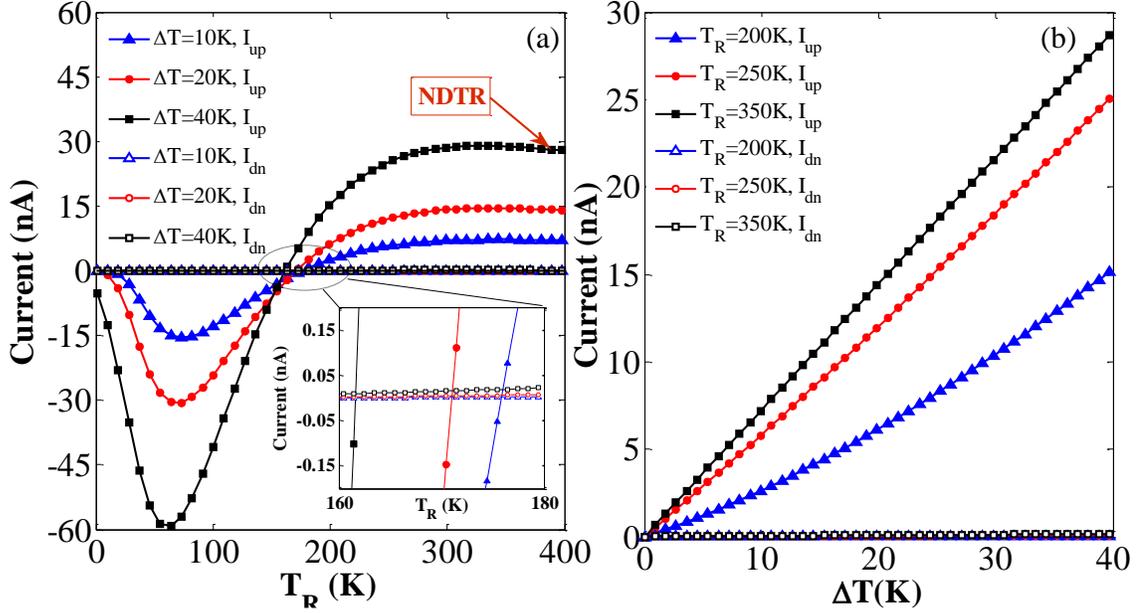

**Figure 5.** (a) The variations of the spin currents versus $T_R$ for $\Delta T$= 10, 20, 40, the spin-up currents ($I_{up}$) have finite values and the spin-down currents ($I_{dn}$) are nearly zero (i.e., SFE), and (b) the spin currents versus $\Delta T$ for $T_R$= 200, 250, and 350 K.

The spin-dependent transmission spectrum is also displayed in figure 6(a). As shown in figure 6(a), the first peak value of the spin-up transmission occurs in the energy values ranging from 0 to 0.03 eV, whereas the second peak value occurs in the energy values ranging from -0.03 to -0.13 eV. Hence, the spin-up dominates the transport properties and yields a nearly perfect SFE. However, the spin-down transmission is almost zero within this range of energies. Like in SDSE case, the combined effects of external fields on the hybrid GSNRs are also evaluated for the SFE case and similar results are obtained. Figure 6(b) and 6(c) shows the band structures of the leads and central region for the considered hybrid GSNRs, respectively. As shown in figure 6(b), the spin-dependent subbands of the leads are almost matched, whereas the lowest-energy subbands of the central region belong to two different spin states (see figure 6(c)). As can be seen from figure 6(c), the ferromagnetic exchange field ($M_z$ = 0.157 eV) causes different spins to shift in opposite directions. The semiconducting gaps are also generated between the hole and electron subbands when a perpendicular and/or transverse electric fields ($E_z$ = 0.030 V/Å, $E_{yS}$ = 0.575 V/Å and $E_{yG}$ = 0.029 V/Å) are applied. However, the transverse electric field causes the electron states to shift to the lower energies, whereas the hole states shift to the higher ones. It is noted that the spin splitting induced by the external electric field is very small and negligible, as shown in inset of figure 6(b). Figures 6(d) to 6(f) show the current spectrum of the spin currents, plotted for different temperature sets. A comparison between the current spectrum of spin-up and spin-down clearly shows that the spin-down spectra are almost zero. However, the spin-up current spectra are dominant in all cases. This again confirms the nearly perfect SFE. For $I_{up}$, the peak value of the current spectrum at $\Delta T$= 20 K is larger than that of $\Delta T$= 10 K when $T_R$ is assumed equal to 350 K. This confirms that the spin current is increased as $\Delta T$ increases. The $J$ area for $T_R$ =300K is smaller than that of $T_R$ =350K when $\Delta T$ is assumed as 20K. This shows that the increase of $T_R$ increases the spin currents.

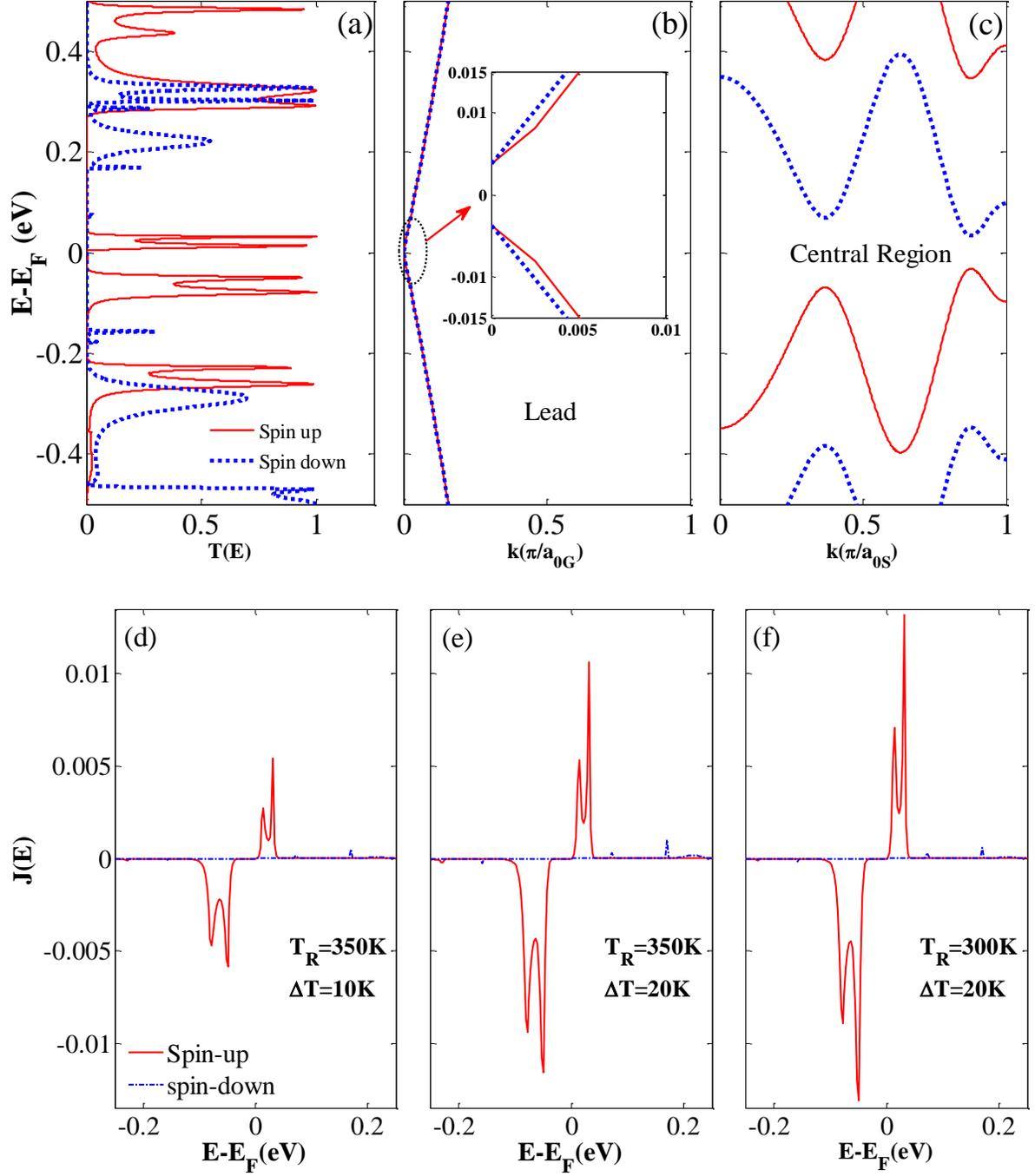

**Figure 6.** (a) The spin-dependent transmission spectra ($T_{up}$ and $T_{dn}$) versus $E-E_F$, (b) the band structures of the right (left) lead with an applied electric field of $E_{yG}= 0.029$ V/Å (inset shows the spin splitting induced by the electric field is very small and negligible) (c) the band structures of the ferromagnetic central region including an exchange field of $M_z= 0.157$ eV, electric fields of $E_Z= 0.030$ V/Å, and $E_{yS}= 0.575$ V/Å, and the variation of the spin-dependent current spectrum $J(E)$ for (d) $T_R = 350$ and $\Delta T = 10$ K, (e) $T_R = 350$ and $\Delta T = 20$ K, and (f) $T_R = 300$ and $\Delta T = 20$ K.

Figure 7 show the variations of spin polarization efficiency, SPE (%) = $(|I_{up}|-|I_{dn}|)/(|I_{up}|+|I_{dn}|) \times 100$, versus $T_R$ and $\Delta T$, respectively. The results clearly show that a high SPE is achieved for the selected $T_R$ values. For example, the SPE is almost equal to 100% for the low temperature values, whereas it is measured about 99% for a wide range of $T_R$ and $\Delta T$ values [12,85]. It is noted that a numerical fluctuation for the $T_R$ values ranging from 140 K to 190 K is also observed. This basically is related to the reverse sign of $I_{up}$ and the mutual competition between $I_{up}$ and $I_{dn}$ in this range of temperatures, as shown in inset of figure 5(a). It is noted that the spin channels of the pristine AGNRs and ZSNRs are partly conductive and the magnitude of the associated SPE is much less than that of the considered hybrid structure.

### 3.3. The effects of central region length

In this subsection, the effects of central region length ($L$) on the thermally-induced current for the considered hybrid ZGSNRs is studied. The variations of the spin currents ($I_{up}$ and $I_{dn}$) versus $T_R$ at $\Delta T$ = 40 K for $M$ = 7–13, and selected external fields has been shown in figure 8(a), in the SDSE case. Indeed, the $L$ parameter proportionally changes with $M$, in all cases. According to figure 8(a), SDSE occurs for $M$ = 8, 9, 12, and 13 for a wide range of $T_R$ values around the room temperature (see the filled darker points). However, $M = 8$ was selected in this research because it can provide a perfect SDSE and stronger spin currents. Figure 8(b) also shows the changes of $I_S$ and $I_C$ versus $M$ at $T_R = 300$ K and $\Delta T = 40$ K. As shown in figure 8(b), $I_S$ is two order of magnitude (i.e., they differ by a factor ranging from 10 to 100) larger than $I_C$ for some $M$ values (see darker filled points). This again confirms that $M = 8$ can suitably provide the largest $I_S$ value and a stronger SDSE for the selected external fields. Figure 8(c) shows the variations of the spin currents versus $T_R$ at $\Delta T = 40$ K for $M = 7$–13 and the adopted external fields, in the SFE case. It is realized that the $I_{up}$ is relatively enhanced for $M = 8$ and leads to a larger thermal-induced spin current and a stronger thermal SFE in the hybrid ZGSNRs. Figure 8(d) also provides evidence that the $I_{up}$ ($I_{dn}$) is two order of magnitude larger than $I_{dn}$ ($I_{up}$) for some $M$ values at $T_R = 300$ K and $\Delta T = 40$ K, in the SFE case (see darker filled points). It is noted that the results reported in this section are all limited to the considered range of parameters and thus they cannot be generalized for various values of external fields.

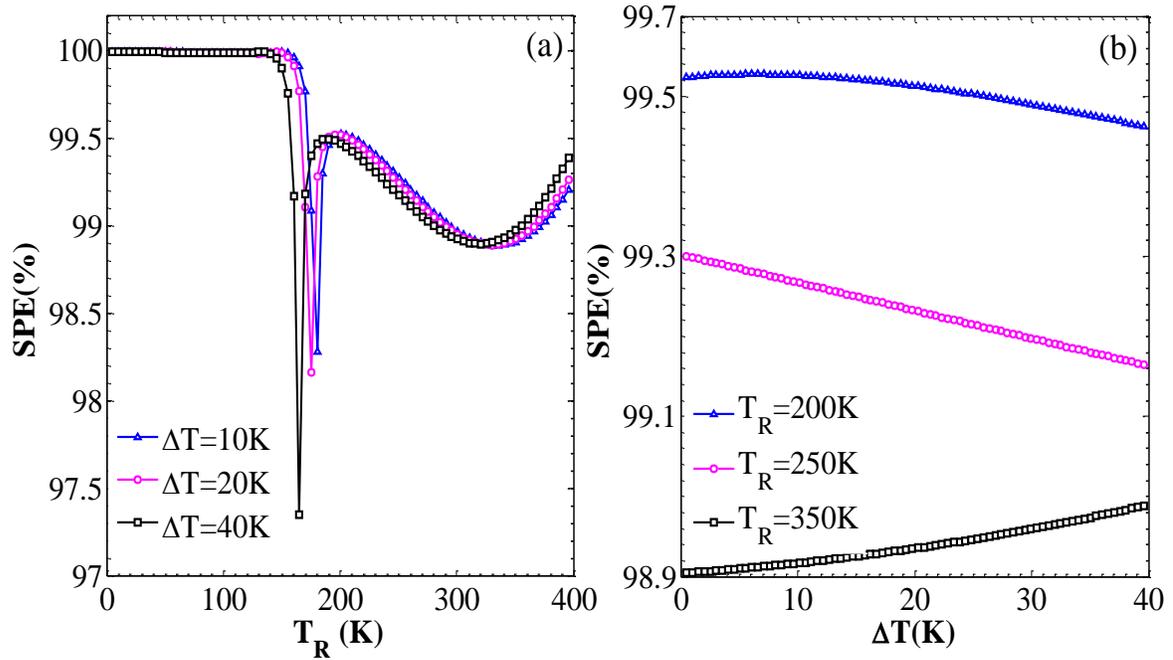

**Figure 7.** (a) The variation of the spin polarization efficiency (SPE) versus $T_R$ for $\Delta T$= 10, 20, and 40 K (b) the SPE variation versus $\Delta T$ for $T_R$ = 200, 250, and 350 K.

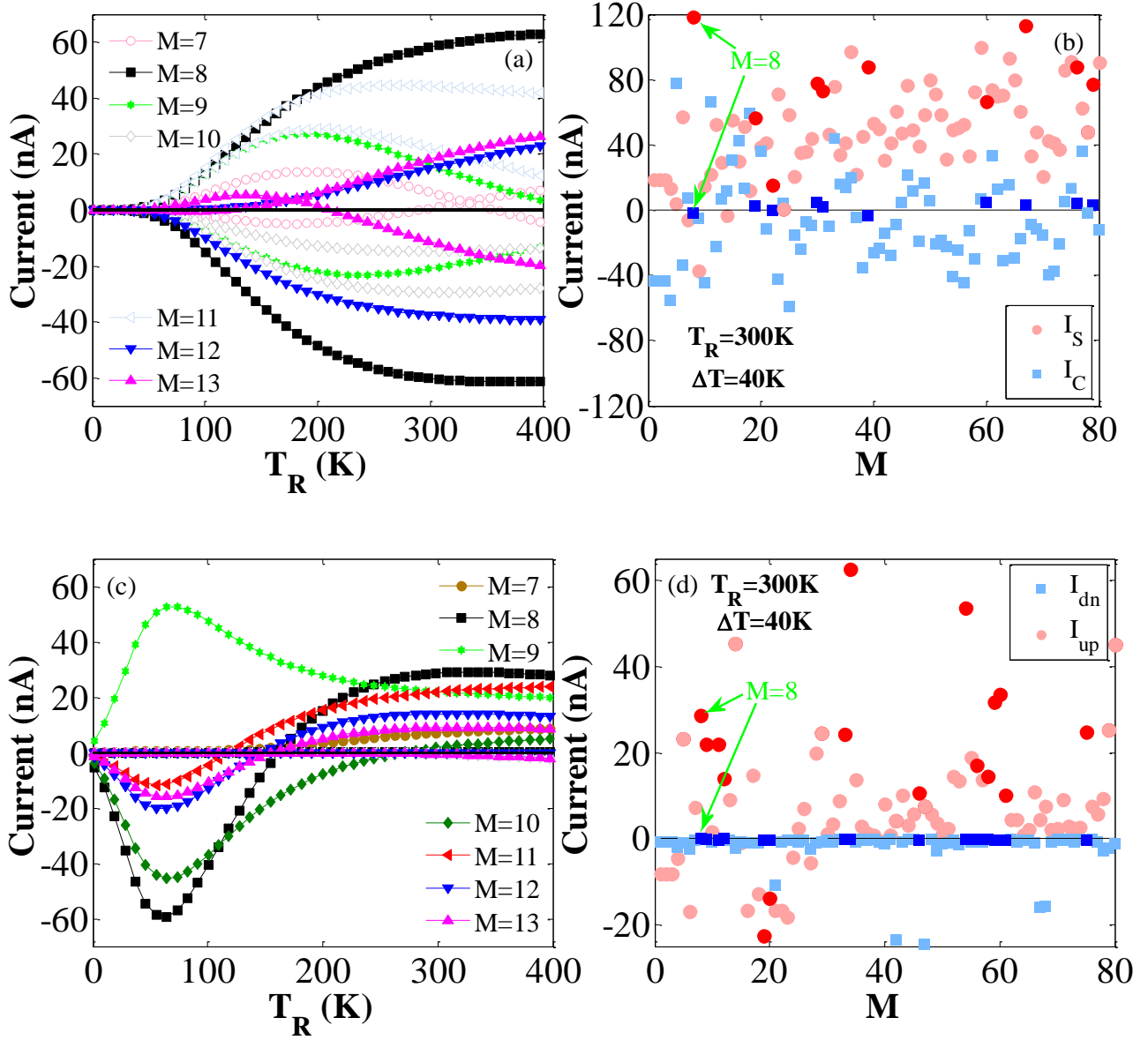

**Figure 8.** The variations of (a) $I_{up}$ and $I_{dn}$ versus $T_R$ at $\Delta T = 40$ K for the lengths of $M = 7$–13; Note that the plots shown in the positive and negative regions of the Y-axis are related to $I_{up}$ and $I_{dn}$, respectively, (b) $I_S$ and $I_C$ versus $M$ at $T_R = 300$ K and $\Delta T = 40$ K, and the selected external fields, in the SSE case, (c) $I_{up}$ and $I_{dn}$ versus $T_R$ at $\Delta T = 40$ K for $M = 7$–13; Note that the plots on the X axis are related to $I_{dn}$, and the others to the $I_{up}$ (d) $I_{up}$ and $I_{dn}$ versus $M$ at $T_R = 300$ K and $\Delta T = 40$ K, and the adopted external fields, in the SFE case.

### 3.4. Thermoelectric performance

The Seebeck coefficients for two different configurations of ferromagnetic exchange and local external electric fields, as discussed above, are studied in this section. Figure 9 shows the variations of the spin-up ($S_{up}$) and spin-down ($S_{dn}$) Seebeck coefficients, namely, the spin ($S_S$) and charge ($S_C$) Seebeck coefficients versus the Fermi energy. The SDSE and SFE are obtained for the hybrid GSNRs for the selected values of exchange and electric fields, as shown in figure 9(a) and 9(b), respectively. In order to balance the thermal forces acting on the charge carriers, a larger bias for the lower electric conductance is often required. This consequently produces larger $S_S$ values [86,87]. Different behaviors are generally observed for the $S_{up}$ and $S_{dn}$ of the two studied configurations. This is especially true for the $E_F$ values ranging from -0.2 to 0.2 eV. In the SFE case, the peak values of -3.67 and 3.77 for the $S_{up}$ and $S_{dn}$ are almost obtained for the $E_F = 0.20$ eV and $E_F = -0.06$ eV values, respectively (see figure 9), whereas these values for the SDSE are equal to 1.73 and -2.0, and occur around the Fermi level. As can be seen from figure 9, when $S_{up}$ is equal to zero for some values of the Fermi energy, $S_{dn}$ is nonzero and vice versa. This is related to the fact that the electron and hole currents are canceled out by each other in one spin channel, whereas there is a thermally-induced spin-polarized current in the another one. For example, according to figure 9(b), in the SFE case, $S_{up}$ values are almost equal to zero for $0.2 < E_F < 0.3$ eV, whereas $S_{dn}$ has nonzero values in this range. It is worth mentioning that $S_{up}$ equals to zero at three different points for the $E_F$ values ranging from -0.2 to 0.2 eV, whereas $S_{dn}$ equals to zero at only one point in this region. As shown in figure 9, $S_{up}$ and $S_{dn}$ also have equal values with different signs at several $E_F$ values. This results in zero $S_C$ and nonzero $S_S$ values in SDSE and SFE cases. It is noted that $S_C$ slightly changes around the $S_C = 0$ line in the SDSE case (see figure 9(a)), whereas in the SFE case, it occurs at some different points (see figure 9(b)). This may be attributed to the fact that the $S_{up}$ and $S_{dn}$ are almost symmetric respect to this line. This shows that $\Delta T$ can produce a pure spin current without any charge current. Hence, a zero-charge voltage and a nonzero spin voltage ($V_S = S_S T$) are generated by the hybrid GSNRs [88]. The transport is created in such an issue by the spin-up holes and spin-down electrons with similar magnitudes but different current directions. Thus, $\Delta T$ can generate a pure spin current and a near-perfect SDSE. It is also observed that $S_S$ is almost flat for the larger values of $E_F$. This may be due to the inversely symmetric relationship of the spin-dependent Seebeck coefficients and their linear dependence on the $E_F$. This leads to a constant difference between the $S_{up}$ and $S_{dn}$. The Seebeck polarization, $P_S = (|S_{up}| - |S_{dn}|) / (|S_{up}| + |S_{dn}|)$ is also shown in figure 9 (see the thicker colored line) to separately specify the effect of each spin channel on the $S_S$. This is performed by changing the color of $S_S$ as the $P_S$ value is changed. As shown in figure 9, because $S_{up}$ ($S_{dn}$) is almost dominant for some values of $E_F$, $S_S$ color tends to magenta (turquoise) color in this region. However, because $S_{up}$ and $S_{dn}$ have almost equal values, but with different signs, the $S_S$ color, almost tends to blue for the entire $E_F$ values. It is worth mentioning that the $S_{up}$ or $S_{dn}$ signs are related to the p- or n-type nature of the device. $S_S$ and $S_C$ have the same sign when the $S_{up}$ is dominant, whereas their signs are opposite as $S_{dn}$ is dominant.

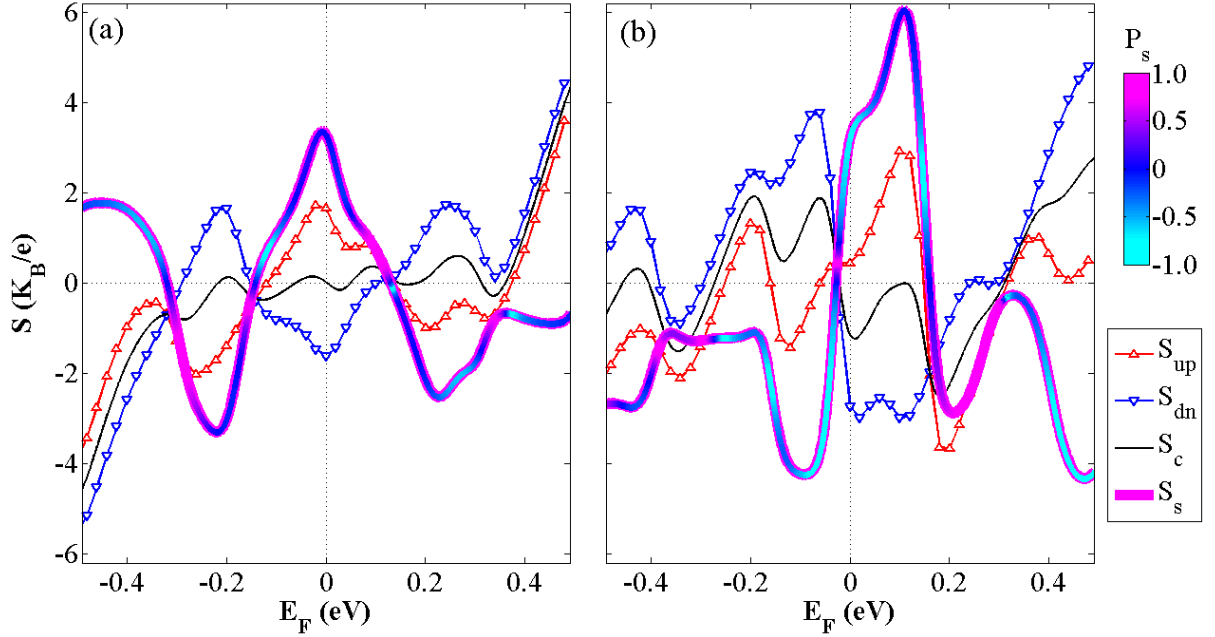

**Figure 9.** The variations of the spin-up ($S_{up}$), spin-down ($S_{dn}$), spin ($S_s$) and charge ($S_c$) Seebeck coefficients versus the Fermi energy and at room temperature for (a) the SSE and (b) the SFE cases. The $S_S$ color changes in accordance with the Seebeck polarization value.

Figure 10 shows the variations of the spin Seebeck ($S_s$) and spin polarization ($P_s$) versus $E_z$ and $M_z$ for different values of the transverse electric fields. Various behaviors can be seen for the $S_s$ and $P_s$ parameters obtained from four different values of the transverse electric fields. As can be seen from the color bars in figure 10, the red and blue colors represent the large and small values for $S_s$ and $P_s$, respectively. $S_s$ and $P_s$ are also considered as odd functions with respect to $M_z$ for the entire plots [89]. This results in $I_S$ direction can be tuned by changing the magnetization direction in the central region. This shows that $I_S$ can magnetically be manipulated. Unlike $E_z$, $S_s$ and $P_s$ can significantly be changed by varying the exchange field. This further illustrates that $S_s$ and $P_s$ can magnetically be employed. It is noted that $S_s$ and $P_s$ are reach their peak values when $M_z$ is almost equal to ±0.1 eV, and with no transvers (or with homogeneous) electric fields. It is noted that these functions are even with respect to the $E_z$ when $E_{yG} = E_{yS} = 0$ (figure 10(a) and 10(b)). The results show that $S_S$ increases when the absolute value of $E_z$ is increased and $E_z < 0$ (see figure 10(c)), whereas the $P_s$ can be large for the entire $E_z$ values. However, the $S_s$ value can almost be reached to ±1 for the larger magnitudes of $E_z$. The maximum value of $S_s$ almost occurs at $M_z=0.1$ eV, as illustrated in figure 10(c). As shown in figures 10, $P_S$ and $S_S$ reach their peak values for a limited number of areas as the inhomogeneous transverse electric fields are adopted. $S_S$ absolute maximum value is increased in SDSE and SFE cases as compared to the two previous cases (see figure 10(e) to 10(h)).

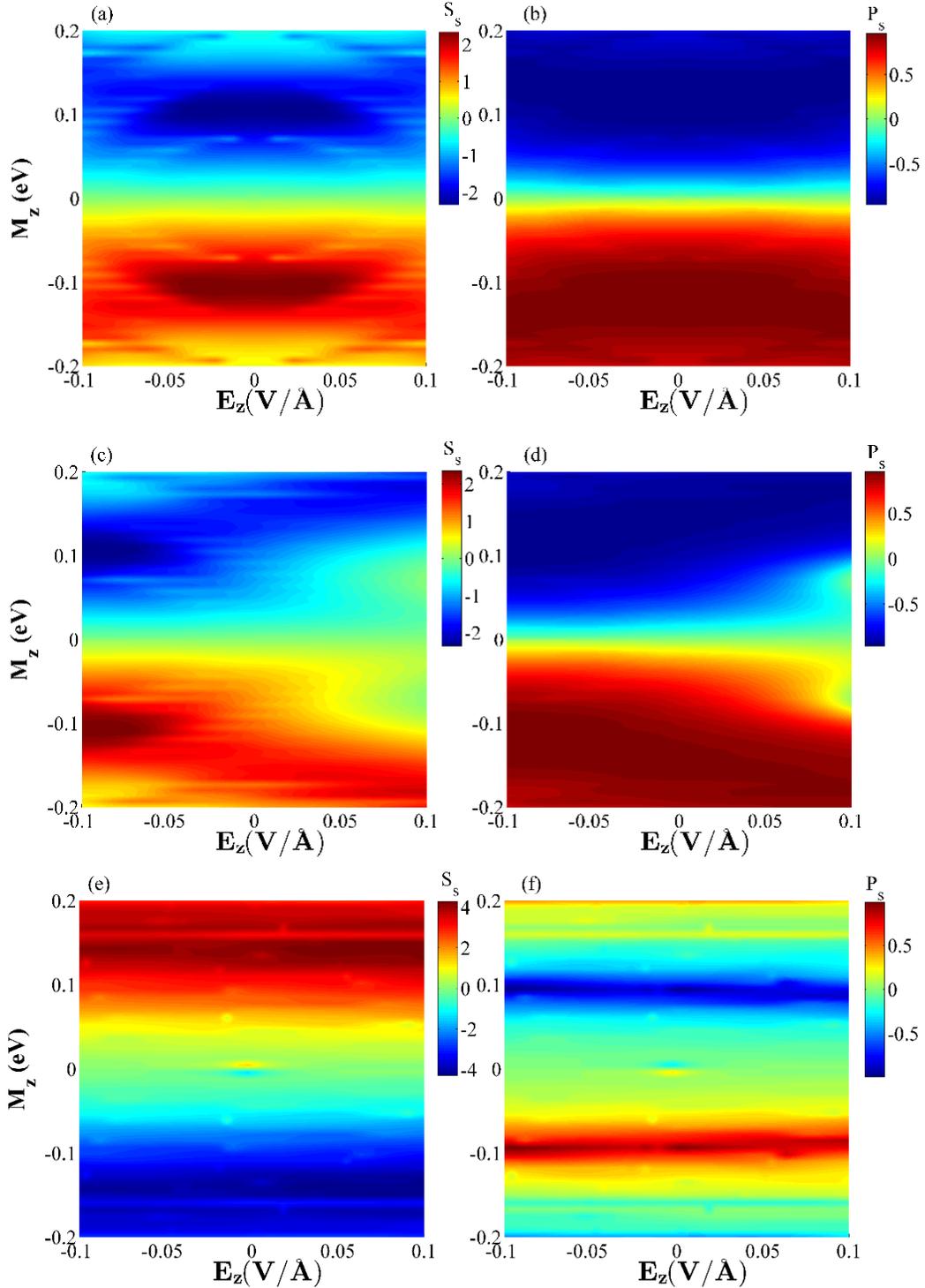

**Figure 10.** The combined effects of ferromagnetic exchange fields, $M_Z$, and external electric fields ($E_Z$, $E_{yG}$ and $E_{yS}$) on the spin Seebeck coefficient ($S_S$) and Seebeck polarization ($P_S$) for the considered hybrid GSN*R* and (a, b) $E_{yG}$ =$E_{yS}$ =0 (c, d) $E_{yG}$ =$E_{yS}$ =0.001 V/Å (e, f) $E_{yG}$ =0.913 V/Å, $E_{yS}$ =0.127 V/Å (g, h) $E_{yG}$ =0.029 V/Å, $E_{yS}$ =0.575 V/Å. The scale of spin Seebeck coefficients in all plots is $k_B/e$.

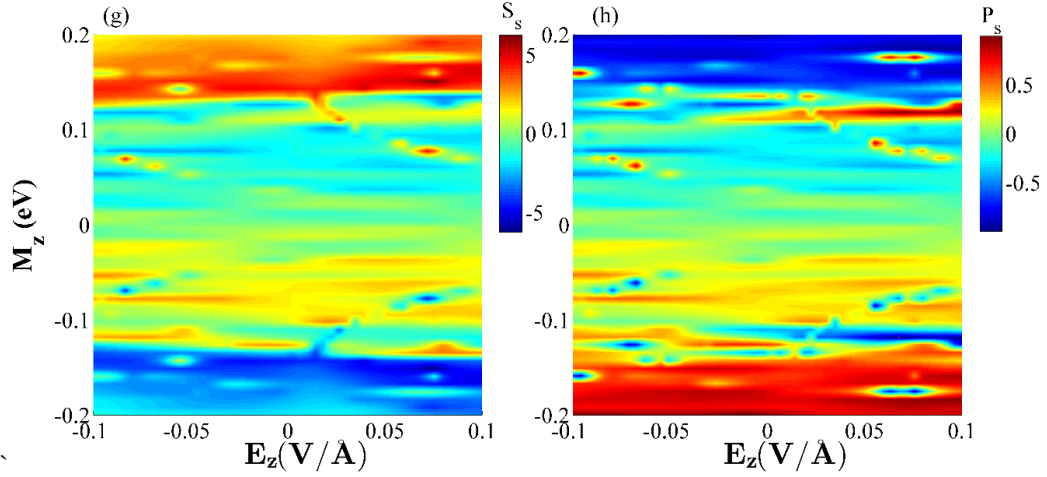

**Figure 10 (Continued).** The combined effects of ferromagnetic exchange fields, $M_Z$, and external electric fields ($E_Z$, $E_{yG}$ and $E_{yS}$) on the spin Seebeck coefficient ($S_S$) and Seebeck polarization ($P_S$) for the considered hybrid GSNR and (a, b) $E_{yG} = E_{yS} = 0$ (c, d) $E_{yG} = E_{yS} = 0.001$ V/Å (e, f) $E_{yG} = 0.913$ V/Å, $E_{yS} = 0.127$ V/Å (g, h) $E_{yG} = 0.029$ V/Å, $E_{yS} = 0.575$ V/Å. The scale of spin Seebeck coefficients in all plots is $k_B/e$.

Figure 11 also shows the variations of the thermal-driven net spin current ($I_S$) and the total charge current ($I_C$) versus $M_z$ and $E_z$ at $T_R$=358 K and $\Delta T$=20 K (see figure 4(c)) for the SDSE case values of the transverse electric fields and in the absence of these fields. Different behaviors can be seen for the $I_S$ and $I_C$ parameters obtained from two plots. As can be seen from the color bars, the red and blue colors represent the large and small values for $I_S$ and $I_C$, respectively. $I_S$ is also considered as an odd function with respect to $M_z$ for the entire plots i.e., $I_S$ sign is varied by changing $M_z$ sign. It is noted that these functions are even with respect to $E_z$ when $E_{yG} = E_{yS} = 0$ (figure 11(a) and (b)). A pure spin current can be observed for a limited range of small and large values of $M_z$ as $I_S$ reaches its peak value, and $I_C = 0$. Figure 11(c) and (d) also provides evidence that a pure spin current can be obtained as the $M_z$ value is almost equal to 0.181 eV (see also figure 4(c)).

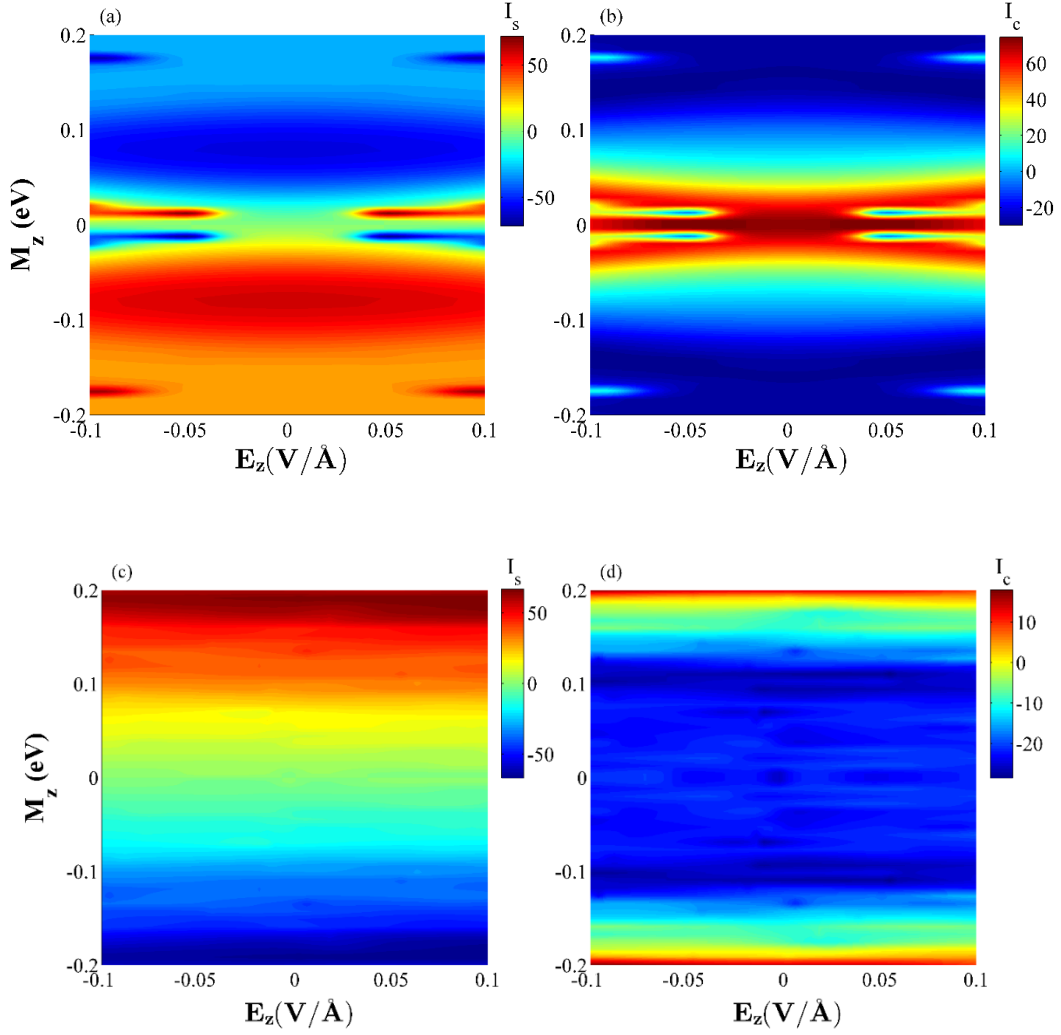

Figure 11. The combined effects of ferromagnetic exchange fields, $M_z$, and external electric fields ($E_z$ and $E_{yG}$, $E_{yS}$) on the net spin current ($I_S$) and total electron current ($I_C$) for the considered hybrid GSNR and (a, b) $E_{yG} = E_{yS} = 0$ (c, d) $E_{yG}$ =0.913 V/Å, $E_{yS}$ =0.127 V/Å. The results are for $T_R$=358 K and $\Delta T$=20 K. $I_S$ and $I_C$ have nA units in all plots.

## 4. Conclusions

To realize the SDSE and the thermal SFE, the spin-dependent thermoelectric transport properties of hybrid GSNRs, as spin caloritronics devices, have been studied in this research. The effects of the temperature gradient between the left and right leads, ferromagnetic exchange fields, $M_z$, and the local external electric fields, $E_z$ and $E_y$, were also included. The results showed that a nearly perfect SDSE could be observed in the hybrid GSNRs. Because the spin-up and spin-down currents are only produced due to the temperature gradient, and they flow in opposite directions with almost equal magnitudes. In the SDSE case, positive spin-up and negative spin-down currents with the threshold temperature $T_{th}$ (~ 50 K) were observed for the values of $M_z$ = 0.181 eV, $E_z$ = 0.081 V/Å, $E_{yS}$ = 0.127 V/Å, and $E_{yG}$ = 0.913 V/Å. Different charge current behaviors have also been obtained by selecting various device temperature sets.

A nearly zero charge thermopower was also obtained, which further demonstrates the emergence of the SDSE. In the SFE case, $M_z = 0.157$ eV, $E_z = 0.030$ V/Å, $E_{yS} = 0.575$ V/Å, and $E_{yG} = 0.029$ V/Å were considered. Larger values of the spin-up current were obtained for the high temperatures, whereas the spin-down current is zero for the entire range of temperatures as the right lead temperature ($T_R$) increases. Thus, a nearly perfect SFE was achieved at room temperature, whereas the spin polarization has reached up to 99%. It is noted that due to the competition between the spin-dependent currents, some interesting transport features such as the change of the flowing direction and negative differential thermoelectric resistance were observed. This evidently confirms the potential thermoelectric device applications of the studied hybrid GSNRs by selecting various device temperature sets. The results reported and discussed in this study are limited to the considered set of parameters and thus they cannot be generalized for various values of external fields.

*Corresponding author: Farhad Khoeini (khoeini@znu.ac.ir)